# Rust and Go Directed Fuzzing with LibAFL-DiFuzz

[1,2] *T.P. Mezhuev, ORCID: 0009-0009-0610-287X <mezhuevtp@ispras.ru>*
[2] *D.A. Parygina, ORCID: 0000-0002-4029-0853 <parygina.darya.2001@yandex.ru>*
[1] *D.O. Kutz, ORCID: 0000-0002-0060-8062 <kutz@ispras.ru>*

[1] *Ivannikov Institute for System Programming of the Russian Academy of Sciences, 25, Alexander Solzhenitsyn st., Moscow, 109004, Russia.*

[2] *Lomonosov Moscow State University, GSP-1, Leninskie Gory, Moscow, 119991, Russia.*

**Abstract.** In modern SSDLC, program analysis and automated testing are essential for minimizing vulnerabilities before software release, with fuzzing being a fast and widely used dynamic testing method. However, traditional coverage-guided fuzzing may be less effective in specific tasks like verifying static analysis reports or reproducing crashes, while directed fuzzing, focusing on targeted program locations using proximity metrics, proves to be more effective. Some of the earliest directed fuzzers are, for example, **AFLGo** and **BEACON**, which use different proximity metric approaches. Although most automated testing tools focus on C/C++ code, the growing popularity of Rust and Go causes the need for precise and efficient testing solutions for these languages. This work expands the applicability of directed fuzzing beyond traditional analysis of C/C++ software. We present a novel approach to directed greybox fuzzing tailored specifically for Rust and Go applications. We introduce advanced preprocessing techniques, rustc compiler customizations, and elaborate graph construction and instrumentation methods to enable effective targeting of specific program locations. Our implemented fuzzing tools, based on LibAFL-DiFuzz backend, demonstrate competitive advantages compared to popular existing fuzzers like afl.rs, cargo-fuzz, and go-fuzz. According to TTE (Time to Exposure) experiments, Rust-LibAFL-DiFuzz outperforms other tools by the best TTE result. Some stability issues can be explained by different mutation approaches. Go-LibAFL-DiFuzz outperforms its opponent by the best and, in the majority of cases, by average result, having two cases with orders of magnitude difference. These results prove better efficiency and accuracy of our approach.

**For citation:** Mezhuev T.P., Parygina D.A., Kutz D.O. Rust and Go directed fuzzing with LibAFL-DiFuzz. Trudy ISP RAN/Proc. ISP RAS, vol. 38, issue 2, 2026, pp. 69-82. DOI: 10.15514/ISPRAS-2026-38(2)-5.

**Acknowledgements:** The results were obtained using the services of the Ivannikov Institute for System Programming (ISP RAS) Data Center.

# Направленный фаззинг Rust и Go с помощью LibAFL-DiFuzz

[1,2] *Т.П. Межуев, ORCID: 0009-0009-0610-287X* <*mezhuevtp@ispras.ru*>
[2] *Д.А. Парыгина, ORCID: 0000-0002-4029-0853* <*parygina.darya.2001@yandex.ru*>
[1] *Д.О. Куц, ORCID: 0000-0002-0060-8062* <*kutz@ispras.ru*>

[1] *Институт системного программирования им. В.П. Иванникова РАН, Россия, 109004, г. Москва, ул. А. Солженицына, д. 25.*

[2] *Московский государственный университет имени М.В. Ломоносова, Россия, 119991, Москва, Ленинские горы, д. 1.*

**Аннотация.** В современном мире при использовании подходов безопасной разработки ПО анализ программ и автоматическое тестирование имеют важное значение для минимизации уязвимостей перед выпуском программного обеспечения, при этом фаззинг является быстрым и широко используемым методом динамического тестирования. Однако традиционный фаззинг с обратной связью по покрытию может быть менее эффективен в некоторых задачах, таких как проверка срабатываний статического анализа или воспроизведение ошибок, где более эффективным оказывается направленный фаззинг, ориентированный на достижение целевых точек программы с использованием метрик близости. Одними из первых направленных фаззеров являются, например, AFLGo и BEACON, которые используют разные подходы метрик близости. Хотя большинство инструментов автоматического тестирования ориентированы на анализ C/C++ кода, растущая популярность Rust и Go вызывает потребность в точных и эффективных решениях для их анализа. Данная работа расширяет применимость направленного фаззинга за пределы традиционного анализа ПО на языках C/C++. Мы представляем новый подход к направленному фаззингу, специально разработанный для приложений на языках Rust и Go. Мы предлагаем новые методы предварительной обработки программы, оптимизации компилятора rustc, а также специальные методы построения графов и инструментации для обеспечения эффективного достижения целевых точек в программе. Реализованные нами инструменты направленного фаззинга для Rust и Go, основанные на инструменте LibAFL-DiFuzz, демонстрируют конкурентные преимущества по сравнению с популярными существующими фаззерами для этих языков, такими как afl.rs, cargo-fuzz и go-fuzz. Согласно экспериментам по метрике TTE (Time to Exposure), Rust-LibAFL-DiFuzz превосходит другие инструменты по наилучшему результату. Некоторые проблемы со стабильностью могут быть объяснены различными подходами к мутации входных данных. Go-LibAFL-DiFuzz превосходит своего оппонента по лучшим показателям и, в большинстве случаев, по среднему результату, имея два случая с разницей в несколько порядков. Эти результаты доказывают более высокую эффективность и точность нашего подхода.

**Ключевые слова:** динамический анализ; фаззинг; направленный фаззинг; инструментация кода; поиск ошибок; цикл разработки безопасного кода; информационная безопасность.

**Для цитирования:** Межуев Т.П., Парыгина Д.А., Куц Д.О. Направленный фаззинг Rust и Go с помощью LibAFL-DiFuzz. Труды ИСП РАН, том 38, вып. 2, 2026 г., стр. 69–82 (на английском языке). DOI: 10.15514/ISPRAS–2026–38(2)–5.

**Благодарности:** Результаты получены с использованием услуг Центра коллективного пользования Института системного программирования им. В.П. Иванникова РАН – ЦКП ИСП РАН.

## 1. Introduction

Program analysis and automated testing have become one of the most critical components in modern SSDLC [1-3]. Proper testing of new software prior to release is necessary to minimize the risk of new vulnerabilities emerging. Among the various testing approaches, fuzzing [4, 5] continues to be one of the most widely used techniques. It is a relatively fast automated dynamic testing method that evaluates program behavior across a broad set of input data. However, for certain specific tasks – such as verifying static analysis reports, testing patches, or reproducing crashes – coverage-guided fuzzing can be less effective, since it guides only for overall coverage. In contrast, directed fuzzing [6] aims at specific program locations (target points, TP) to achieve these objectives faster. This approach uses specialized proximity metrics to measure how close a given program execution is to reaching the designated target points.

While the great majority of the automated testing tools are restricted with C/C++ code analysis, these languages are not the only ones that need to be checked. For instance, there is a vast range of critical components written in compiled languages such as Rust and Go. Nowadays, these programming languages are becoming more and more popular for developing applications. Along with this, the demand for reliable testing continues to grow. To achieve maximum benefits, various testing methods should be applied, including fuzzing and directed fuzzing.

In this paper, we make the following contributions:

- We propose an approach to enable directed fuzzing of Rust and Go applications. We add generic functions handling at preprocessing stage, *rustc* compiler customization for Rust directed fuzzing. We implement a combined approach to graph construction with correct debug information based on AST and SSA IR for Go code. We also propose our feedback instrumentors that work on the source code of Go programs.
- We implement the first Rust and Go directed fuzzing tools based on LibAFL-DiFuzz [7] backend. We evaluate our tools in comparison with afl.rs [8, 9], cargo-fuzz [10], and go-fuzz [11] tools using Time to Exposure metric. According to TTE experiments, Rust-LibAFL-DiFuzz outperforms other tools by the best result. Some stability issues can be explained by different mutation approaches. Go-LibAFL-DiFuzz outperforms its opponent by the best and, in the majority of cases, by average result, having two cases with orders of magnitude difference. These results prove better efficiency and accuracy of our approach.

## 2. Related work

Although there are several fuzzing tools for Rust and Go programs, none of them implement directed approach. In this section, we describe the most popular tools for Rust and Go languages [8-14] as well as some examples of directed fuzzers for C/C++ languages [6, 15-17] to demonstrate the difference between a coverage-guided and directed approaches.

### 2.1 afl.rs

Relying on effectiveness of AFL++, afl.rs [8, 9] was created as its adaptation to Rust language analysis. It aims to generate inputs that trigger panic errors and memory corruptions. The tool is integrated with cargo, and can be used inside Rust ecosystem.

The tool leverages the capacity of Rust compiler to generate LLVM IR [18]. Similarly, to AFL++, afl.rs provides compiler wrapper that instruments the target program on the basic block level using special Rust compiler flags. Code instrumentation provides coverage collection, branch hits tracking, and interaction with the fuzzer.

For better usability, afl.rs provides fuzz! macro allowing to write the test harness as a closure. The macro then reads bytes from the standard input stream and redirects it to the closure. The fuzzer has a special panic processing mechanism that ensures catching the panic and calling process::abort instead to report error correctly.

### 2.2 cargo-fuzz

Cargo-fuzz [10] is another fuzzer adaptation to Rust language. Under the hood it invokes libFuzzer [4] to perform analysis. All the stages are also integrated to cargo [19] ecosystem.

Cargo-fuzz utilizes LLVM Sanitizer Coverage [20] instrumentation to retrieve the code coverage information. LibFuzzer is integrated with the help of libfuzzer-sys [21] wrapper library. All fuzzing targets need to be saved in a separate crate that depends on the target program.

Cargo-fuzz supports fuzz_target! macro to write the test harness inside. For this harness, libFuzzer is repeatedly invoked with new generated inputs. The fuzzing stops as an error occurs.

## 2.3 CrabSandwich

Another fuzzing tool for Rust language is based on a Rust fuzzing library. CrabSandwich [12] works on LibAFL [22] backend instead of libFuzzer, extending cargo-fuzz set of supported tools. The main goal of the tool's authors was to minimize user efforts for creating new fuzzers and migrating existing ones to the new backend. Thus, the CrabSandwich user interface closely follows the one of cargo-fuzz.

Technically, CrabSandwich consists of two crates: the runtime and the linking component. Runtime is based on LibAFL library utilizing its mechanisms for receiving program execution feedback, setting up execution strategies, and applying custom mutators. Like libFuzzer, CrabSandwich runtime performs in-process target execution, and implements input rejection strategy. Unlike libFuzzer, it applies effective scheduling strategies from AFL++ [5], and adds some new mutations: conditional Grimoire, comparison interceptors and string-oriented mutations, comparison log.

The second crate is the replacement for libfuzzer-sys crate used by cargo-fuzz. Its purpose is to link the harness with the CrabSandwich runtime with minimal efforts for migration from cargo-fuzz. In general, the crate provides thin user interface to the actual runtime.

## 2.4 go-fuzz

The most popular fuzzing tool for Go is go-fuzz [11]. Unlike Rust fuzzers, go-fuzz does not just adapt existing fuzzing tools to Go program analysis. It is a separate complete tool written in Go itself and included in a Go toolchain.

Go-fuzz takes inspiration from AFL++ fuzzing approach, implementing similar techniques for coverage-guided analysis. The program preparation stage however is different. The tool instruments the target program on source code level. Following libFuzzer style, go-fuzz needs a special Go-wrapper and performs in-process fuzzing. It also supports libFuzzer engine integration.

Go-fuzz uses return code values to evaluate and classify generated inputs. Inputs that discover new coverage are saved for further use; the other ones are ignored. The tool also supports Go modules analysis.

## 2.5 GoLibAFL

Another tool to benefit from LibAFL features is GoLibAFL [13, 14]. This tool aims to combine support for latest Go versions and analysis flexibility. One of the important LibAFL features used by GoLibAFL is comparison tracing support.

GoLibAFL uses Sanitizer Coverage [20] instrumentation supported by both Go toolchain and LibAFL. The actual fuzzer is written in Rust code based on LibAFL library applying in-process fuzzing executor. The ability to interact with Go target program is achieved by Foreign Function Interface (FFI). It is the bridge between the two languages based on a C-compatible interface. The harness for GoLibAFL is written in Go and is exported as a libFuzzer-style harness function.

## 2.6 AFLGo

One of the earliest approaches to directed greybox fuzzing is the AFLGo tool [6]. It reaches specified target points in the code by calculating input energy using a *simulated annealing algorithm* [23, 24]. Within this fuzzer, inputs that lead closer to the target points receive higher energy values.

AFLGo formulates target exploration as an optimization task. Over time, inputs receiving the highest energy values asymptotically converge toward a set of globally optimal solutions, corresponding to execution paths that minimize the distance to the target locations. The convergence speed is controlled through an exponentially decreasing temperature parameter, which determines the probability of receiving high energy to suboptimal solutions.

## 2.7 BEACON

BEACON [15] employs *reachability analysis* to eliminate irrelevant basic blocks. First, it calculates the reachability of each block to the target point and prunes those that are irrelevant, so they are not instrumented for metrics feedback. The remaining blocks are used in taint analysis to apply constraints on program variables. Such approach significantly reduces overhead but restricts BEACON to single-target scenarios only.

In the next phase, *backward interval analysis* is applied. The program is sliced based on statically computed control flow, and value intervals are computed for all variables in the remaining basic blocks. These intervals are used to generate target reachability formulas, which are inserted in instrumentation as runtime checks. If the constraints are satisfied, execution proceeds toward the target; otherwise, it terminates.

# *3. LibAFL-DiFuzz directed fuzzing approach*

The directed fuzzing approach that we use in our LibAFL-DiFuzz tool is possible to be implemented, generally speaking, for any language that supports C-code function calls. In order to prepare the target for directed fuzzing, four steps must be performed: 1) static analysis to obtain enhanced target sequences (ETS) – the lists of basic blocks leading to target points with computed context weights; 2) target program instrumentation based on the ETS got on previous step, i.e., implementation of *ETS feedback*; 3) *SanCov* instrumentation for *coverage feedback*; 4) linking with *libforkserver* library, containing all required utilities for fuzzing. A complete description of this directed fuzzing approach has been previously given in our articles [7, 25]. In this article we propose LibAFL-DiFuzz directed fuzzing approach for Rust and Go languages.

Fig. 1 shows a generalized pipeline for any language on how to build a target for directed fuzzing using LibAFL-DiFuzz.

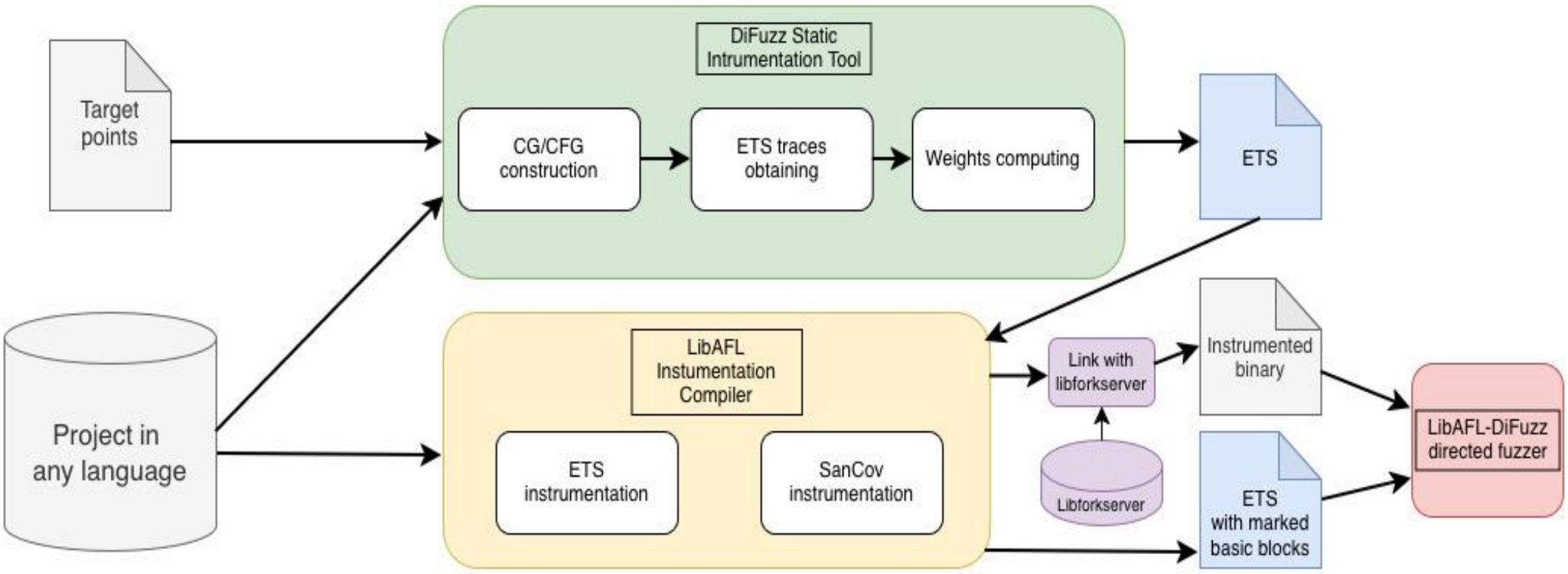

*Fig. 1. Fuzz target preparation for LibAFL-DiFuzz.*

It should be noted here that ETS instrumentation inserts code that writes the target basic blocks unique IDs to *ETS shared memory*, so that an ETS trace can be collected at runtime. A call to *libafl_start_forkserver()* is inserted at the beginning of the program to configure its interaction with the LibAFL-DiFuzz.

S*anCov* instrumentation, in turn, inserts calls to *__sanitizer_cov_trace_pc_guard()* into all possible basic blocks to provide coverage guidance, as well as calls to *__afl_map_shm()* and *__sanitizer_cov_trace_pc_guard_init()* at the beginning of the program, which configure *coverage shared memory*.

And to make it work together, we link the target binary to our *libforkserver* library, that contains binary code of all functions mentioned above and symbols that are used for *ETS shared memory* interaction.

## *4. Rust directed fuzzing*

Building fuzzing targets for Rust is quite similar to how it is done for C/C++. Since Rust compilation is based on the LLVM toolchain [18] (which we use in LibAFL-DiFuzz), preparing Rust targets becomes conceptually similar to C/C++: we construct call graphs (CGs) and control flow graphs (CFGs) from LLVM (in DOT format), analyze them with DiFuzz static preprocessing tool, and then instrument the binary file using special LLVM-passes.

Fuzzing Rust code requires modification to two components. First, we adapted DiFuzz static preprocessing tool (hereinafter referred as DiFuzz-Rust for Rust) to handle Rust's specific features. Second, we modified the *rustc* compiler itself to instrument the code.

### 4.1. DiFuzz-Rust: static analysis

Following the same approach as for C/C++ targets, LLVM IR is used for static analysis of Rust code. The *DiFuzz-LLVM* library [7] is used to build a CGs and CFGs with the necessary debug information about functions and basic blocks, such as source code positions, implicit calls, etc. DiFuzz-Rust then analyzes these graphs and builds ETS from them. Finally, the program is instrumented based on those ETS. One of the main changes in DiFuzz-Rust (and also in DiFuzz) is support for template functions processing. Previously, if there were several functions with identical code positions in the call graph, which in most cases means that it is a template function, the control flow graph was constructed and analyzed for only one randomly selected function. Now, for all of such functions ETS is marked, and then instrumentation is added to each of them in the binary code, not just to one of them.

### 4.2. Binary instrumentation with libafl-rustc

It was decided to modify the *rustc* compiler to instrument Rust code. We add default compilation options so that LLVM-passes for ETS and *SanCov* are used automatically during the the target building. Also, any code optimizations are disabled. The customized compiler was called *libafl_rustc*.

Thus, the Rust project instrumentation pipeline is conceptually almost identical to the C/C++ pipeline, with the main differences being only in the technical implementation.

## *5. Go directed fuzzing*

The approach to support Go projects significantly differs from how it was implemented for C/C++ and Rust. The key difference is that Go compilation is based on its own intermediate representation (IR) called *SSA IR* [26], instead of LLVM toolchain. Therefore, we implemented a separate *DiFuzz-Go* library to build graphs and obtain debug information for functions and basic blocks. For target instrumentation step, we created tools that implement ETS and *SanCov* feedback insertion, similar to LLVM-passes.

### 5.1. DiFuzz-Go: static analysis

First of all, it is required to implement the *DiFuzz-Go* library to construct call graphs and control flow graphs in the same format (DOT) used for other supported languages (C/C++/Rust). Since this graph format is not native to Go *SSA IR*, we had to build graphs with the required debug information using internal graph construction libraries and then convert them to DOT using our formatter. After that, in general, it is already possible to perform static analysis with the DiFuzz tool for Go projects, obtaining the same ETS at the output.

The listings on Fig. 2-4 show the difference between DOT graph nodes and edges for C/C++, Rust and Go.

```
Node0x7f2ac4ce0410[shape=record,filename="/xlnt/source/styles/format.cpp",startline=177,endline=179,
bbendline=177,startcolumn=5,label="{_ZNK4xlnt6format12fill_appliedEv}"];
Node0x7f2ac4ce0410 -> Node0x7f2ad405c1b0;
Node0x7f2ac4ce0410 -> Node0x7f2ad40567c0[indirect];
```

*Fig. 2. DOT graph node, edge and indirect edge for C/C++.*

```
Node0x7ff148049ed0[shape=record,filename="/goblin/src/archive/mod.rs",startline=181,endline=189,
bbendline=182,startcolumn=0,label="{_ZN6goblin7archive6Member13extended_name17hcab95c3126c53047E}"];
Node0x7ff148049ed0 -> Node0x7ff14801f2a0;
Node0x7ff148049ed0 -> Node0x7ff14804e260[indirect];
```

*Fig. 3. DOT graph node, edge and indirect edge for Rust.*

```
Node0x503d486ed8f2[shape=record,filename="/fyne/widget entry.go",startline=1950,endline=1969,bbendline=1951,
startcolumn=0,label="{fyne.io/fyne/v2/widget.isWordSeparator}"];
Node0x503d486ed8f2 -> Node0xc8fad8770c5f;
Node0x503d486ed8f2 -> Node0x77fc110bd1dc[indirect];
```

*Fig. 4. DOT graph node, edge and indirect edge for Go.*

### 5.2. Go instrumentors

The key difference for Go instrumentation is that our instructions are inserted at the source code level, rather than at the intermediate representation level (as in C/C++, Rust). It is not low-level instructions that are inserted into the code, but their semantically equivalent high-level commands. Therefore, our instrumentation tool must do the following:

1) take the target source code and ETS information as input;

2) search for the target basic blocks;

3) instrument them and produce modified source code with added commands.

Moreover, we need to create another tool that implements *SanCov* instrumentation for *coverage feedback*, which is done automatically by *libafl_cc/cxx* compilers for C/C++ and by *libafl-rustc* compiler for Rust (with providing several additional options). It works as follows: the tool goes through each basic block of every program function and inserts *SancovGuard* function call that represents LibAFL's coverage feedback.

As a result, two frontend instruments were created to build Go targets. They insert high-level commands and function calls into the program's source code, thereby implementing *ETS* and *coverage feedbacks* necessary for the LibAFL-DiFuzz directed fuzzer. Fig.5 shows how such instrumentation looks like at the source code level for Go.

## *6. Implementation*

This part describes the implementation details and difficulties for each language. For Rust it is about determining the program binary format for easier analysis with DiFuzz-Rust, and the process of creation a modified compiler. For Go we had to decide which instrumentation approach is the best to use, which libraries and tools are most suitable, and how to use them to obtain the information required for analysis and implement instrumentation that is semantically similar to LLVM IR approach.

### 6.1. DiFuzz-Rust: difference from DiFuzz for C/C++

Although *rustc* uses LLVM for compilation, the output code has certain features and patterns that should be taken into account when building CGs and CFGs. First, the DiFuzz-Rust tool needs to get LLVM bytecode file as input. It can be obtained for Rust by passing the "*--emit=llvm-bc*" option to *rustc* compiler and then combining all "*.bc*" modules into a single file. Second, inline instructions (*DILocation* LLVM IR nodes with the parameter "*inlinedAt = %i*") are often inserted into functions, which need to be "unrolled" to get the real position in source file. Finally, different *rustc* compilation options sometimes produce different debug information about the column numbers in the source file, so it was decided to ignore all information about columns for Rust.

```go
func check_label() {
outer:
for i := 0; i < 3; i++ {
    for j := 0; j < 3; j++ {
        if i == j {
            fmt.Println("label")
            break outer
            }
        }
    }
    fmt.Println("done")
}
```

```go
func check_label() {
    SancovGuard(21)
outer:
    SancovGuard(22)
    for i := 0; i < 3; i++ {
        SancovGuard(23)
        InstrumentETS(2497)
        for j := 0; j < 3; j++ {
            SancovGuard(24)
            InstrumentETS(7023)
            if i == j {
                SancovGuard(25)
                InstrumentETS(1110)
                fmt.Println("label")
                break outer
            }
        }
    }
    SancovGuard(22)
    fmt.Println("done")
}
```

*Fig. 5. Example of Go high-level instrumentation.*

## 6.2. libafl-rustc compiler

It was decided to compile *rustc* from its source code, since the stable build updates the LLVM version with each new release, and we only support LLVM-18 for analysis. Also, building *libafl_rustc* from source allows to configure the compiler so that the arguments required for instrumentation are automatically applied when used, as well as arrange the compilation so that all dependencies are loaded together with the compiler.

The following *rustc* arguments are necessary to perform instrumentation. “*-C opt-level=0*” is required to be passed to disable optimization, “*-Z /path/to/difuzz-ets-pass.so -C passes=difuzz-ets-pass*” to enable the LLVM-pass that performs *ETS feedback* instrumentation, "*-C passes=sancov-module -C llvm-args=-sanitizer-coverage-level=3 -C llvm-args=-sanitizer-coverage-trace-pc-guard*" to enable *SanCov* instrumentation for *coverage feedback*, and, finally, "*-L/path/to/libforkserver -lforkserver*" to link with *libforkserver*. This ensures correct target instrumentation, and the fuzzing target could be built using the *cargo* builder by specifying the path to the modified *libafl-rustc* compiler in the "*RUSTC*" environment variable.

Full pipeline for Rust target building for LibAFL-DiFuzz looks like in Fig. 6.

## 6.3. DiFuzz-Go: difference from DiFuzz for C/C++

The DiFuzz-Go library was implemented to obtain call graphs and control flow graphs. The library is based on the native Go libraries “*go/ast*”, “*go/ssa*”, “*go/parser*”, “*go/packages*”, “*go/cfg*”, and “*go/callgraph*”. A distinctive feature of the Go analysis approach is that these libraries allow to skip the code compilation step and simply take the path to the program source code directory as input.

The *SSA IR* of the program can be used to build a CG. But building a CFG for a function requires the program's *AST*, since the library for constructing these graphs, “*go/cfg*,” requires a high-level code representation. This leads to the main difficulty: *SSA IR* and *AST* often have different debug information about function start and end lines. In general, this leads to inaccurate ETS analysis. Therefore, it was decided to use a combined approach to CG constructing: first, a call graph is constructed using *SSA IR*, then CFGs are constructed for each function in the graph, which are used to collect unified debug information. Furthermore, when the target functions are known during static analysis, their control flow graphs are already constructed with the same debug information, which increases the accuracy of ETS.

Once the CG or CFG with debug information has been constructed, it must be converted to our modified DOT format, compatible with DiFuzz tool. To do this, a special formatter traverses the graph in depth and sequentially prints the nodes (with debug information), and their edges. In addition, implicit calls are not analyzed since Go does not have OOP as such.

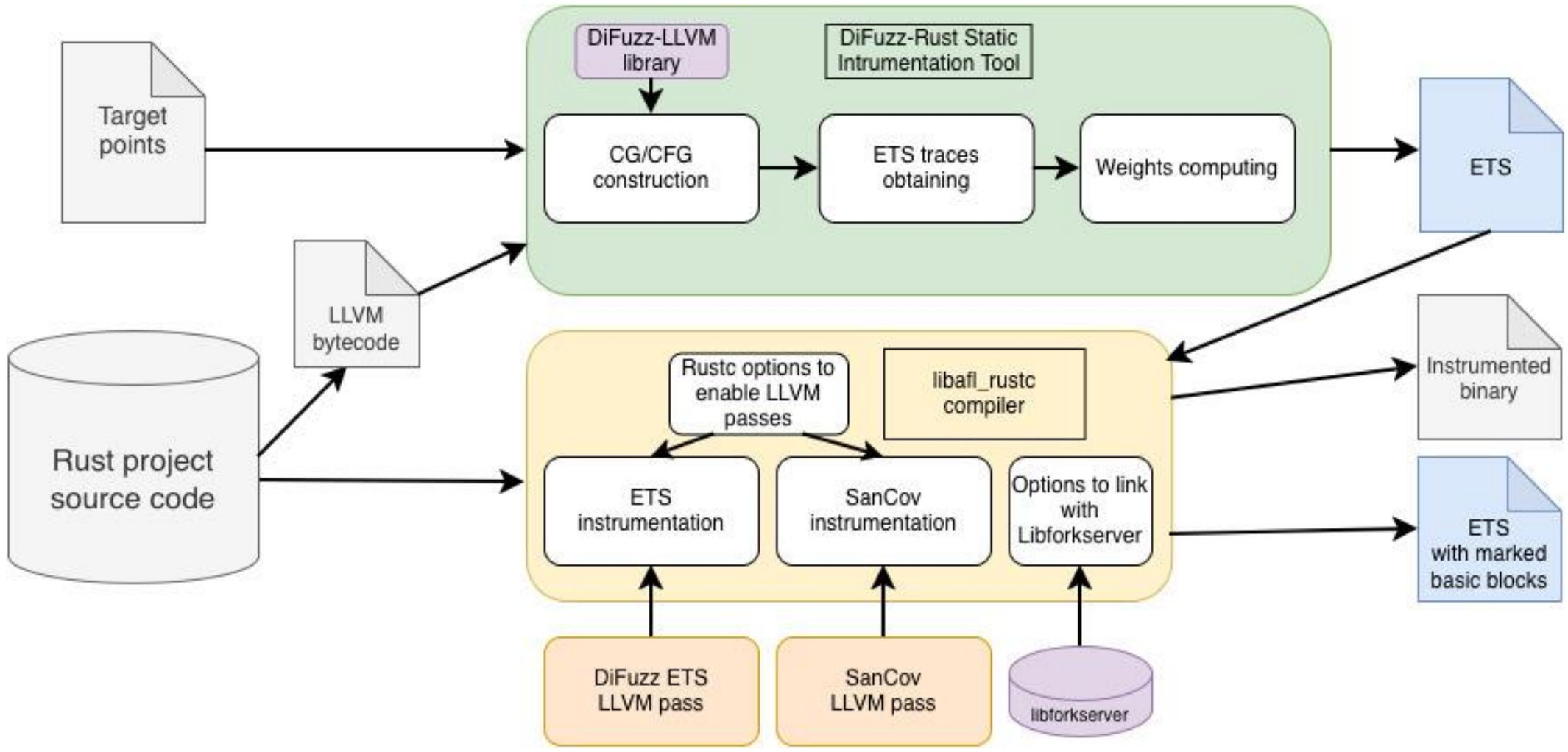

*Fig. 6. Fuzz target pipeline preparation for Rust.*

## 6.4. Go intrumentors

As mentioned earlier, ETS and *SanCov* instrumentation is performed at the source code level. This is because “*go/ssa*” does not allow modification of *SSA IR*, and using non-native tools for that is not entirely reliable, so it was decided to implement instrumentation using the “*go/ast*” library that works with *AST*.

The main idea is to add the *DiFuzz-Instr* library to the project source code, which contains the instrumenting functions *InstrumentETS()* and *SancovGuard()*, whose calls will be inserted into source code. *InstrumentETS()* contains code that sends visited basic block ID to the fuzzer’s *ETS shared memory*. The *SancovGuard()* function is the “bridge” that calls the C function *__sanitizer_cov_trace_pc_guard()* internally. Using the capabilities of the Go language, the *DiFuzz-Instr* library also contains two *init()* functions that call *libafl_start_forkserver()* and *__afl_map_shm()* with *__sanitizer_cov_trace_pc_guard_init()* for ETS and *SanCov*, respectively. This init-mechanism in Go allows to avoid inserting these calls into the source code, since all *init()* functions are called before the *main* function is entered.

To describe the instrumentation stage, let's start with ETS. The tool is given the path to the directory with the target source code and the *ets.toml* file containing information about ETS. Next, the tool searches all source files (specified in *ets.toml*) for the target basic blocks:

1) An *AST* for the current file is constructed;
2) All functions in it are iterated through;

3) CFG is constructed for each function, and all basic blocks are traversed in search of those blocks that are present in ETS;
4) Each basic block found is assigned a unique ID, and then the *AST* tree is searched for an expression that is located inside this basic block;
5) If such an expression is found, a call to the *IntsrumentETS(ID)* function must be inserted before it, specifying the block ID in its parameters; *DiFuzz-Instr* library is imported to the file that contains this basic block.

*SanCov* instrumentation is performed in a similar way and is even simpler. We need to provide the path to the directory with the target source code. Then the instrument goes through all possible "*.go*" files and tries to insert a call to *SancovGuard(ID)* into each basic block, where *ID* is the unique basic block identifier.

*libforkserver* can be linked to the target with passing CGO_LDFLAGS="-L/path/to/libforkserver -lforkserver" environment variable to *go build* command.

The result is the following pipeline for preparing a Go target for LibAFL-DiFuzz:

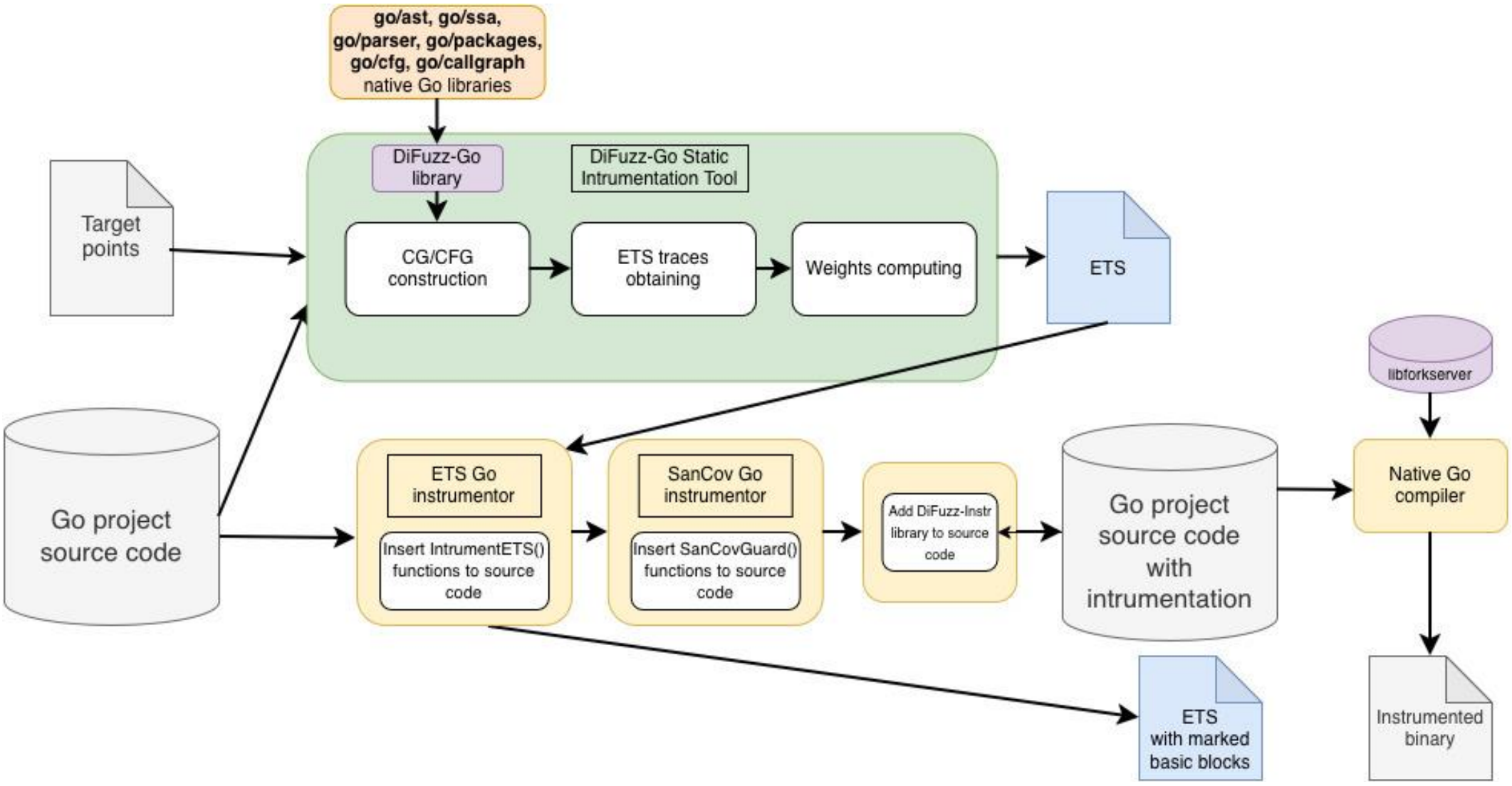

*Fig. 7. Fuzz target pipeline preparation for Go.*

## 7. Evaluation

We evaluated our approach to Rust and Go directed fuzzing using TTE (Time to Exposure) metric that shows the time spent on crash discovering. Experimental setup included one machine with two 64-Core AMD EPYC 7702 CPUs, 256 GB of RAM, and Ubuntu 20.04 LTS, and one machine with two 32-Core AMD EPYC 7542 CPUs, 512 GB of RAM, and Ubuntu 20.04 LTS. We compared our directed fuzzer Rust-LibAFL-DiFuzz with afl.rs and cargo-fuzz. We compared our directed fuzzer Go-LibAFL-DiFuzz with go-fuzz tool. For the evaluation, we selected three projects for Rust: goblin, gdb-command, libhtp-rs; and three projects for Go: ollama, image-go, fyne. For each project, we added some synthetic errors (with the help of *panic()* calls) to some code points. The points of *panic()* calls insertion are the target points for analysis. The set of chosen target points with their IDs, locations, and execution timeouts, is shown in Table 1.

Each experiment was run several times with a predefined timeout. For each execution, we measured time intervals between the experiment start and different target points discovering. After the end of experiment, we checked crash seeds with *gdb* to validate the *panic()* call is reached. Experiments results for the best attempts and average TTE values are shown in Tables 2 and 3. Columns represent different instruments, lines – different target points, and each cell contains TTE value in seconds.

*Table 1. Target points chosen for experiments.*

| TP ID | Project | Location | TimeOut, s |
|---|---|---|---|
| $goblin_1$ | goblin | src/mach/mod.rs:187 | 60 |
| gdb-$command_1$ | gdb-command | src/stacktrace.rs:157 | 1800 |
| gdb-$command_2$ | | src/mappings.rs:114 | |
| libhtp-$rs_1$ | libhtp-rs | src/test.rs:252 | 60 |
| $ollama_1$ | ollama | harmony/harmonyparser.go:221 | 300 |
| $ollama_2$ | | parser/parse.go:416 | |
| image-$go_1$ | image-go | jpeg/reader.go:330 | 1800 |
| image-$go_2$ | | png/reader.go:174 | |
| image-$go_3$ | | tiff/reader.go:393 | |
| image-$go_4$ | | webp/decode.go:175 | |
| $fyne_1$ | fyne | storage/repository/parse.go:91 | 1800 |
| $fyne_2$ | | internal/cache/svg.go:24 | |

*Table 2. TTE experiments results for Rust projects.*

| TP ID | Rust-LibAFL-DiFuzz | | afl.rs | | cargo-fuzz | |
|---|---|---|---|---|---|---|
| | best, s | avg, s | best, s | avg, s | best, s | avg, s |
| $goblin_1$ | **1.9** | 8.3 | 3.0 | 25.4 (20% TO) | 3.2 | **6.2** |
| gdb-$command_1$ | **14.0** | 241.7 (13% TO) | 23.0 | **70.2** | 35.0 | 86.0 |
| gdb-$command_2$ | **1.0** | 366.1 (13% TO) | 3.0 | **4.0** | 198.0 | 240.8 |
| libhtp-$rs_1$ | **2.3** | 44.0 | 3.0 | **4.0** | 23.0 | TO (90% TO) |

*Table 3. TTE experiments results for Go projects.*

| TP ID | Go-LibAFL-DiFuzz | | go-fuzz | |
|---|---|---|---|---|
| | best, s | avg, s | best, s | avg, s |
| $ollama_1$ | **0.5** | 7.9 | **0.5** | **1.7** |
| $ollama_2$ | 9.0 | **9.5** (50% TO) | **2.0** | 48.0 (20% TO) |
| image-$go_1$ | **0.4** | **1.88** | 1.0 | 1.9 |
| image-$go_2$ | **0.4** | 2.2 (50% TO) | 1.0 | **1.3** |
| image-$go_3$ | **0.5** | **16.9** (20% TO) | 745.0 | 1153.7 (70% TO) |
| image-$go_4$ | **0.6** | **2.1** | 1183.0 | 1406.8 (60% TO) |
| $fyne_1$ | **1.9** | 10.5 | 3.0 | **4.5** |
| $fyne_2$ | **4.4** | **11.3** | 6.0 | 13.7 |

Some cells contain percent of timeout executions – it means that the instrument was not able to reach target point within predefined timeout.

TTE results show that Rust-LibAFL-DiFuzz beats other fuzzers by the best result on all programs. In best try, our tool shows better results than afl.rs and much better than cargo-fuzz. Compared to cargo-fuzz, Rust-LibAFL-DiFuzz reached the target point in much less time, especially on the *gdb-command* program (almost 200 times faster in one case). This proves Rust-LibAFL-DiFuzz approach efficiency for reaching target points within a small amount of time. However, our tool underperforms others by average TTE value, that pinpoints some stability issues. On the mentioned *gdb-command* program, our tool couldn't reach target points faster on average than even cargo-fuzz – mostly due to a lot of timeouted runs. That means the fuzzer got stuck and can't explore program to reach a specified target point. For example, we couldn't even measure an average value for cargo-fuzz on *libhtp-rs* because of that (90% of all runs were timeouted). By average TTE time, almost on all programs afl.rs has the best result. Generally speaking, such results can be explained by different mutation approaches of the tools that can lead to efficiency variations in different programs. For instance, AFL has the fixed set of mutation operators that are applied to the input during two strictly defined stages. Such strategy may cause aggressive input transformation that is good for discovering new program paths. LibAFL offers flexible mutation pipeline with customizable operators set and configurable order, which in some cases can lead to insufficient input transformation. The issue of determining the best mutation set and order for directed fuzzing is a topic for further research.

As for Go-LibAFL-DiFuzz, it also outperforms its opponents by the best TTE result in almost all cases. Furthermore, in the majority of cases average results are also better for our tool, and the average timeout percent is lower than for go-fuzz. Special attention should be paid to $image\text{-}go_3$ and $image\text{-}go_4$ target points, where Go-LibAFL-DiFuzz outperforms go-fuzz by orders of magnitude. The only case ($ollama_2$) where our tool loses in the best TTE try, has a better average TTE result than go-fuzz: it was able to reach target point 5 times faster despite having 50% timeout rate. Overall, Go-LibAFL-DiFuzz have better both best and average TTE on 4 cases from 8. On the rest cases our tool has better best TTE or average TTE: there are no programs where Go-LibAFL-DiFuzz underperforms go-fuzz on both metrics. These results mean better efficiency and stability of our approach. The only cases where Go-LibAFL-DiFuzz lost to go-fuzz by average TTE not because of timeouts is $ollama_1$ and $fyne_1$. It also could be explained by different mutation strategies used in go-fuzz (which implements AFL++ approach).

## 8. Conlclusion

In this paper, we propose an approach to enable Rust and Go applications testing with directed fuzzing method. We modify preprocessing method for the two languages in different ways. For Rust we add special generic functions handling in program preprocessing stage, and customize rustc compiler. For Go we propose a combined approach to graph construction with correct debug information based on AST and SSA IR. We also propose coverage and ETS frontend instrumentors that work on the source code of Go programs.

We implement the first Rust and Go directed fuzzing tools based on LibAFL-DiFuzz backend. We evaluate our tools in comparison with afl.rs, cargo-fuzz, and go-fuzz tools using Time to Exposure metric. According to TTE experiments, Rust-LibAFL-DiFuzz outperforms other tools by the best TTE result. Some stability issues can be explained by different mutation approaches. Go-LibAFL-DiFuzz outperforms its opponent by the best and, in the majority of cases, by average result, having two cases with orders of magnitude difference. These results prove better efficiency and accuracy of our approach.

## *Информация об авторах / Information about authors*

Тимофей Павлович МЕЖУЕВ – магистр Московского государственного университета имени М.В. Ломоносова, старший лаборант Института системного программирования. Сфера научных интересов: символьное выполнение, гибридный фаззинг, направленный фаззинг.

Timofey Pavlovich MEZHUEV – master of Lomonosov Moscow State University, senior laborant of the Institute for System Programming of the RAS. Research interests: symbolic execution, hybrid fuzzing, directed fuzzing.

Дарья Алексеевна ПАРЫГИНА – аспирант Московского государственного университета имени М.В. Ломоносова. Сфера научных интересов: символьное выполнение, гибридный фаззинг, направленный фаззинг.

Darya Alekseevna PARYGINA – postgraduate student at Lomonosov Moscow State University. Research interests: symbolic execution, hybrid fuzzing, directed fuzzing.

Даниил Олегович КУЦ – кандидат технических наук, младший научный сотрудник Института системного программирования. Сфера научных интересов: динамический анализ, фаззинг, символьное выполнение, гибридный фаззинг.

Daniil Olegovich KUTZ – Cand. Sci. (Tech.), junior research assistant of the Institute for System Programming of the RAS. Research interests: dynamic analysis, fuzzing, symbolic execution, hybrid fuzzing.